\documentstyle[12pt,epsf]{article}
      \newcommand{\beq}{\begin{equation}}
      \newcommand{\eeq}{\end{equation}}
      \newcommand{\beqa}{\begin{eqnarray}}
      \newcommand{\eeqa}{\end{eqnarray}}

      \newcommand{\Tr}{{\rm Tr}}
      
      \newcommand{\bra}{\left\langle}
      \newcommand{\ket}{\right\rangle}
      
      \newcommand{\del}{\partial}

      \newcommand{\al}{\alpha}
      \newcommand{\be}{\beta}

\begin{document}
\begin{titlepage}
\begin{flushright}
NUP-A-97-5 \\
cond-mat/9702156\\
February, 1997
\end{flushright}
\begin{center}
{\large{\bf DIAGRAMMATIC METHOD FOR WIDE CORRELATORS
IN GAUSSIAN ORTHOGONAL AND SYMPLECTIC RANDOM MATRIX
ENSEMBLES}} \\[12mm]
%
%
{\sc Chigak Itoi},
\footnote{\tt e-mail: itoi@phys.cst.nihon-u.ac.jp} \\[2mm]

and 

{\sc Yoshinori Sakamoto}
\footnote{\tt e-mail: yossi@phys.cst.nihon-u.ac.jp} \\[3mm]
{\it Department of Physics, 
College of Science and Technology, Nihon University, 
Kanda Surugadai, Chiyoda-ku, Tokyo 101, Japan} \\[6mm]

{\bf Abstract}\\[5mm]
{\parbox{13cm}{\hspace{5mm}
We calculate connected correlators in time dependent Gaussian orthogonal
and symplectic random matrix ensembles by a diagrammatic method.
  We obtain averaged one-point Green's functions in the leading order
$O(N^0)$ and wide two-level and three-level correlators in the first nontrivial order by summing over twisted and untwisted planer diagrams. \\[3mm]
}}
\end{center}
\vspace*{5ex}
\end{titlepage}

\paragraph{Introduction}
Recently, Br{\'e}zin and Zee pointed out the universality of 
wide connected correlators in random matrix theories  
\cite{BZ}. They recognized
its importance from the viewpoint of physics in disordered systems,
though this universality in 
random matrix theories for two dimensional quantum gravity
was already shown by Ambj{\o}rn, Jurkiewicz and Makeenko \cite{AJM}.
This universal feature of the wide connected correlator is 
shown by explicit calculations in quite extensive classes 
of random matrix ensembles
in various ways \cite{B,BHZ,AA}, while 
the well-known short distance correlator 
can be calculated in limited narrow classes \cite{BZ,mehta}.
We have to take into account 
these purely mathematical properties of
random matrix theories themselves in order to recognize 
the essentially universal nature of level statistics
of physical systems.

In this letter, we  calculate 
the wide connected two level and three level correlators 
in gaussian orthogonal (GOE) and symplectic (GSE) ensembles
with time dependence as discussed in some unitary ensembles
\cite{BZ2,BH,B2,SA}. 
The diagrammatic method is known to be useful to
calculate wide connected correlators in some unitary 
invariant ensembles \cite{BZ2,BZ3} and in GOE \cite{VWZ}. 
We are going to calculate the explicit form of the two level and three level
correlator in GOE and GSE by this method. The results in GOE 
is consistent with that already obtained by Verbaarschot,  Weidenm{\"u}ller and  Zirnbauer. We review their method for correlators in 
GOE and extend it to those in GSE.
For these ensembles, we have to sum over both 
orientable and non-orientable diagrams.
The results can be checked to compare those 
calculated by solving functional equations \cite{B,I} or by replica
method \cite{IMS}, when time dependence is suppressed.  

\paragraph{Gaussian orthogonal ensemble}
To begin with, we explore the case of the GOE which is defined by 
the ensemble of real symmetric matrices  obeying the  probability distribution \begin{equation}
        P(H) = \frac{1}{Z}  \exp \left( -\frac{N}{2} \int_{-\infty} ^\infty
         dt \ \Tr
 \ [\left( \frac{dH}{dt} \right)^2 + m^2 H^2 ] \right), \ \ 
\label{prob} 
\eeq
where $H$ is an $N \times N$ matrix for GOE.  
The measure $D H$ is explicitly written as 
\beq
        D H = \prod_{k = 1}^N dH_{k\, k} \prod_{i<j} dH_{i\,j}.
\label{omeasure}  
\eeq
A dressed line is defined by
\beq
        G_{ij}(z) \equiv \langle \,
        \left(\frac{1}{z - H(0)} \right)_{ij} \,
         \rangle
         \equiv \int DH \,P(H) 
        \left( \frac{1}{z - H(t)}\right)_{ij}, 
\label{G(z)}  
\eeq
which is independent of time $t$ 
because of the transrational invariance of 
the distribution function $P(H)$.
The averaged one-point Green's function $G(z)$
is calculated from this one-point function $G_{ij}(z)$
\begin{equation}
G(z) \equiv \frac{1}{N} \langle \Tr \frac{1}{z-H} \rangle 
= \frac{1}{N} \sum_{i=1} ^N G_{ii}(z).
\end{equation}
The one-point function $G_{ij}(z)$ 
is expanded in the power series of $H$
\begin{equation}
G_{ij}(z) = \frac{1}{z} \sum_{n=0} ^\infty \frac{1}{z^n}\langle 
\left( H^n \right)_{ij} \rangle
\label{PS1}
\end{equation}
We evaluate this function with the diagrammatic decomposition
by using the free propagator of the real symmetric matrix
\begin{equation}
\langle H_{ij}(s) H_{kl}(t) \rangle = 
\frac{1}{2N} ( \delta_{il} \delta_{jk} + \delta_{ik} \delta_{jl} ) K(s-t),
\end{equation} 
where $K(t) = \frac{1}{2m} e^{- m|t|}$.
we can extend the time dependence of the probability distribution
as the time dependent factor $K(t)= \sigma^2 e^{-u(t)}$
with an arbitrary suitable function $u(t)$.
The corresponding diagram to this free propagator 
which has a twisted part is shown
in fig 1. 
The one-point function can be calculated by summing over 
all oriented planer diagrams in eq(\ref{PS1}). 
Any diagram with twisted propagators does not contribute 
to the leading order in the $1/N$ expansion,
and therefore 
 the same diagrams contribute
to the leading order as in the unitary ensemble.
When the series eq(\ref{PS1}) is written in
\begin{equation}
G_{ij}(z)= \frac{\delta_{ij}}{z} 
\sum _{n=0} ^\infty \left( \frac{\sigma^2}{2 z^2}\right)^{n} g_n,
\end{equation}
$g_n$ represents the number of the oriented 
planer diagrams with $n$ free propagators.
The number
$g_n$ obeys the following recursion relation
\begin{equation}
g_{n+1} = g_n + \sum_{m=0} ^{n-1} g_m g_{n-m}, \ \ \ \ n > 1,
 \label{recurs}
 \end{equation}  
 where $g_0 = 1$. 
The corresponding diagrammatic representation 
to the recursion relation (\ref{recurs}) is depicted by fig 2.
On the other hand, the vacuum polarization function 
\begin{equation}
\Sigma(z) \equiv z - G(z)^{-1}
= z \sum_{n=1} ^\infty s_n \left( \frac{\sigma^2}{2 z^2} \right)^n.
\end{equation}
satisfies the equation
\begin{equation}
G(z)(z-\Sigma(z))=1.
\end{equation}
This gives the relation among the expansion coefficients $g_n$ and $s_n$
\begin{equation}
\sum_{m=0} ^n s_n g_{n-m} = 0, \ \ \ \ n > 1,
\label{rel}
\end{equation}
where the coefficient $s_0 \equiv -1$. 
We can show the following relation from the eq(\ref{rel})
by the mathematical inductivity
\begin{equation}
s_n = g_{n+1} - \sum _{m=0} ^{n-1} g_m g_{n-m}.
\label{sn}
\end{equation}
This equation indicates the vacuum polarization function
consists of the one-particle irreducible diagrams.
This equation (\ref{sn})
and the recursion relation (\ref{recurs}) give 
$$
s_n = g_n, \ \ \ \ n > 1,
$$
and therefore
\begin{equation}
\Sigma(z) = \frac{\sigma^2}{2} G(z) = \frac{\sigma^2} {2(z-\Sigma(z))}
\label{vac}
\end{equation}
The vacuum polarization function $\Sigma(z)$
is obtained by solving the quadratic equation 
$$
\Sigma(z) = \frac{1}{2}(z-\sqrt{z^2 - 2 \sigma^2}).
$$
Then we obtain the averaged one-point Green's
function
\beq
        G(z) = \frac{1}{\sigma^2}( z - \sqrt{z^2 - 2 \sigma^2}). 
\label{GOE1}
\eeq
The obtained function  $G(z)$ satisfies 
 the boundary condition $G(z) \rightarrow 1/z$ as 
$z \rightarrow \infty$.

Now we turn to the connected two level correlator 
\beq
        G(z_1, \,  z_2; \ t) \equiv \bra \frac{1}{N} \Tr \frac{1}{z_1 - H(t)} \, 
        \frac{1}{N} \Tr \frac{1}{z_2 - H(0)} \ket_c .
\label{G2} 
\eeq
We calculate this function in the following 
expression of the resolvent operator where we have to only sum 
over simpler diagrams following Br{\'e}zin and Zee \cite{BZ3}
\begin{equation}
G(z_1, \ z_2 ; \ t) = 
\frac{1}{N^2} \frac{\partial}{\partial z_1} \frac{\partial}{\partial z_2}
\sum_{m=1} ^\infty  \sum_{n=1} ^\infty \frac{1}{m n} \frac{1}{z_1 ^m z_2 ^n}
\langle  \Tr H(t)^m 
 \Tr H(0)^n  \rangle_c .
\end{equation}
First, we ignore contractions within the same trace, in which
we take into account only $m=n$ parts
\begin{equation}
\sum_{n=1} ^\infty
\frac{1}{n^2 (z_1 z_2)^n} \langle \Tr H(t)^n \Tr H(0)^n \rangle_c.
\end{equation} 
In the $n$th order, $n$ untwisted planer diagrams and
$n$ twisted planer 
diagrams in fig 3 contribute to the leading order
\begin{equation}
\frac{1}{N^2} \frac{\partial^2}{\partial z_1 \partial z_2}
\sum_{n=1} ^\infty \frac{2n}{n^2} 
\left( \frac{ K(t)}{ 2 z_1 z_2} \right)^n=
- \frac{2}{N^2} \frac{\partial^2}{\partial z_1 \partial z_2}
 \log \left(1- \frac{K(t)}{2 z_1 z_2}  \right).
 \label{tranc}
\end{equation}
Next we include contractions within the same trace in 
$\langle \Tr H(t)^m \Tr H(0)^n \rangle$.
The following operator product expansion (OPE) formula
\begin{equation}
\langle \, \Tr \log \left( 1 - H(t)/z_1 \right) H_{ij}(0) \, \rangle
= 2 \frac{K(t)}{2}  \langle \, 
\left(\frac{1}{z_1-H(t)} \right)_{ij} \, \rangle ,
\end{equation}
tells us that the two level correlator is obtained by
replacing the free one point Green's function $\delta_{ij}/z$
to the dressed one $G_{ij}(z)$ in eq(\ref{tranc})
\begin{equation}
G(z_1, \ z_2 ; \  t) = - \frac{2}{N^2} \frac{\partial^2}{\partial z_1 \partial z_2}
\log \left( 1 - \frac{K(t)}{2}  G(z_1) G(z_2)  \right)
\label{G2GOE}
\end{equation}
This result with respect to the connected part agrees with that obtained by 
solving functional equations \cite{B,I} and by a replica method \cite{IMS}.

\paragraph{Gaussian symplectic ensemble}
The Gaussian symplectic ensemble(GSE) is the ensemble  of 
 quaternion real Hermitian matrices. 
The component $H_{i j} ^{\alpha \beta}$ of $2N \times 2N$ 
quotation real Hermitian matrix $H$
is written as 
\begin{equation}
	H_{i j} ^{\alpha \beta}
	= H^{(0)}_{i j} \delta^{\alpha \beta} 
	+ i \sum_{a = 1}^3 H^{(a)}_{i j} 
                    	\sigma_k ^{\alpha \beta},
\end{equation}
where $\sigma_k ^{\alpha \beta}$ $(k=1, 2, 3 ; \ 
\alpha, \beta = 1, 2)$ is a component of the $2 \times 2$ Pauli matrix,
$H^{(0)}_{i j}$ is a real symmetric matrix and  
$H^{(k)}_{i j}$  $(i, j =1, \cdots, N ) $ is real antisymmetric matrix
in $H_{ij}^{\al\be}$. 
The trace of $H^2$ in eq.(\ref{prob}) reads 
\beq
	\Tr H^2 = \sum_{ij \alpha \beta} 
	{H_{i j}^{\al \be}}{H_{ji}^{\be \al}}
	= 2 \sum_{i=1} ^N  \sum_{j=1} ^N  \sum_{a=0} ^3  ( H^{(a)} _{ij} )^2.
\eeq
Here we define the distribution function of GSE 
\begin{equation}
        P(H) = \frac{1}{Z}  \exp \left( -\frac{N}{4} \int_{-\infty} ^\infty
         dt  \Tr
 \ [\left( \frac{dH}{dt} \right)^2 + m^2 H^2 ] \right), \ \ 
\label{prob} 
\eeq
with the measure   
\beq
	D H  = \prod_{i \leq j} d H_{ij}^{(0)}
	\prod_{k = 1}^3  \prod_{i < j} d H_{ij}^{(k)}. 
\eeq
The free propagator of $H$ is 
\begin{equation}
\langle H_{ij} ^{\alpha \beta} (s) H_{kl}^{\gamma \epsilon}(t) \rangle
= \frac{1}{N} K(s-t) \left( \delta_{il} \delta_{jk} 
\delta^{\alpha \epsilon} \delta^{\beta \gamma}
+\delta_{ik} \delta_{jl} A^{\alpha \beta, \gamma \epsilon} \right),
\end{equation}
where the tensor $A^{\alpha \beta, \gamma \epsilon}$ is defined by
\begin{equation} 
A^{\alpha \beta, \gamma \epsilon}
\equiv \delta^{\alpha \beta} \delta^{\gamma \epsilon}
-\delta^{\alpha \epsilon} \delta^{\beta \gamma}.
\label{deftensor}
\end{equation}
The first and second terms in the free propagator are represented
by an untwisted and twisted diagrams 
shown in fig 4 (a) and (b), respectively.
The calculation of the dressed line
is done by the expansion 
\begin{equation}
G_{ij} ^{\alpha \beta}(z) = 
\langle \left(\frac{1}{z - H(0)}\right)_{ij} ^{\alpha \beta}
\rangle =
\frac{1}{z} \sum_{n=0} ^\infty \frac{1}{z^n} \langle 
\left( H^n \right)_{ij} ^{\alpha \beta} \rangle
\label{PS2}
\end{equation}
The one-point Green's function is defined as its trace
\begin{equation}
G(z) = \frac{1}{2N} \langle \Tr \frac{1}{z-H(0)} \rangle =
\frac{1}{2N} \sum_{i \alpha} 
G_{ii} ^{\alpha \alpha} (z).
\end{equation}
As in the GOE case, only untwisted planer diagrams contribute the 
one-point Green's function in the large $N$ leading order
\begin{equation}
G_{ij} ^{\alpha \beta}(z)= \frac{\delta_{ij} \delta^{\alpha \beta}}{z} 
\sum _{n=0} ^\infty \left( \frac{2 \sigma^2}{z^2}\right)^{n} g_n. 
\end{equation}
Since the number $g_n$ of the oriented 
planer diagrams with $n$ free propagators is identical to 
that in the GOE case, the vacuum polarization function $\Sigma(z)$
is calculated by summing over one-particle irreducible diagrams.
\begin{eqnarray}
\Sigma(z) &\equiv& z - G(z)^{-1} \nonumber \\
&=& 2 \sigma^2 G(z) = \frac{2 \sigma^2} {(z-\Sigma(z))}.
\label{vacGSE}
\end{eqnarray}
The vacuum polarization function $\Sigma(z)$
is obtained by solving the quadratic equation 
$$
\Sigma(z)= \frac{1}{2}(z-\sqrt{z^2 - 8 \sigma^2}).
$$
Then we obtain the averaged one-point Green's
function
\beq
	G(z) = \frac{1}{4 \sigma^2} 
	\left( z - \sqrt{z^2 - 8 \sigma^2} \right).  
\label{GSE1}
\eeq
Note that $G(z)$ tends to $1/z$ as 
$z \rightarrow \infty$.
  
Next we compute the two-point function $G(z_1, z_2)$ defined by  
\begin{eqnarray}
G(z_1,\ z_2 ; \ t) &\equiv& \bra \frac{1}{2 N} \Tr \frac{1}{z_1 - H(t)} \, 
\frac{1}{2 N} \Tr \frac{1}{z_2 - H(0)} \ket_c \nonumber \\
&=& \frac{1}{4 N^2} 
\frac{\partial}{\partial z_1} \frac{\partial}{\partial z_2} 
\sum_{m=1} ^\infty  \sum_{n=1} ^\infty \frac{1}{m n} 
 \frac{1}{z_1 ^m z_2 ^n} \langle  \Tr
H(t)^m \
\Tr H(0)^n  \rangle_c .
\end{eqnarray}
Ignoring contractions within the same trace in the following 
expression
\begin{equation}
\sum_{n=1} ^\infty
\end{equation}
It is the same as in the GOE case 
that in the $n$th order $n$ untwisted planer diagrams and
$n$ twisted planer 
diagrams in fig 3(a)(b) contribute to the leading order.
A part of the twisted diagram shown in fig 3(c) is calculated by
the following contraction formula for the tensor 
$A^{\alpha \beta, \gamma \epsilon}$
defined by eq(\ref{deftensor})
\begin{eqnarray}
(A^n)^{\alpha_1 \alpha_{n+1}, \beta_1 \beta_{n+1}} &\equiv& 
\sum_{\alpha_2, \cdots, \alpha_{n}, \beta_2, \cdots, \beta_{n}}
A^{\alpha_1 \alpha_2, \beta_1 \beta_2} \ A^{\alpha_2 \alpha_3, \beta_2 \beta_3}
\ \cdots \ A^{\alpha_{n} \alpha_{n+1}, \beta_{n} \beta_{n+1}}  \nonumber \\
&=& 2^{n-1} A^{\alpha_1 \alpha_{n+1}, \beta_1 \beta_{n+1}}.
\label{contraction}
\end{eqnarray}
This gives a trace formula
\begin{equation}
\Tr A^n \equiv \sum_{\alpha, \beta} (A^n)^{\alpha \alpha, \beta \beta}  
= 2^n
\end{equation}
which enables us to obtain the two level correlator in this truncation 
\begin{equation}
\frac{1}{4 N^2} \frac{\partial^2}{\partial z_1 \partial z_2}
\sum_{n=1} ^\infty ~ \frac{n}{n^2}~ 
\left( \frac{K(t) }{z_1 z_2} \right)^n (2^n + \Tr A^n) =
-\frac{1}{2 N^2} \frac{\partial^2}{\partial z_1 \partial z_2}
 \log \left( 1- \frac{2 K(t)}{z_1 z_2}  \right).
 \label{trancGSE}
\end{equation}
>From this expression, 
we obtain the exact two level correlator by
replacing the free line $\delta_{ij} \delta^{\alpha \beta}/z$
to the dressed one $G_{ij} ^{\alpha \beta}(z)$ in eq(\ref{trancGSE})
to incorporate contractions within the same trace
\beq
	G(z_1,\ z_2 ; \ t ) = - \frac{1}{2 N^2} 
	\frac{\del^2}{\del z_1 
	\del z_2} 
	\log \left(1 - 2  K(t) G(z_1) G(z_2) \right).
\label{GSE2}  
\eeq
This result with respect to the connected part agrees with that obtained by 
solving functional equations \cite{B,I} and by replica method \cite{IMS}.

\paragraph{Three level correlator}
We can also calculate the connected 
three level correlator both in GOE and GSE.
That in the GOE case is defined by  
\begin{eqnarray}
& & G(z_1, \, z_2, \, z_3 ; t_1, \, t_2, \, t_3)
\equiv  \langle  \, \frac{1}{N}  \Tr \frac{1}{z_1-H(t_1)} \,
\frac{1}{N} \Tr \frac{1}{z_2-H(t_2)} \,
\frac{1}{N} \Tr \frac{1}{z_3-H(t_3)} \, \rangle_c \nonumber \\
&=& - \frac{1}{N^3} \frac{\partial^3}{\partial z_1 \partial z_2 \partial z_3}
\sum_{n_1 = 1} ^\infty \sum_{n_2=1} ^\infty \sum_{n_3=1} ^\infty
\frac{1}{n_1 n_2 n_3} \langle \, \Tr 
\left( \frac{H(t_1)}{z_1} \right)^{n_1}
\Tr \left(\frac{H(t_2)}{z_2} \right)^{n_2} 
\Tr \left( \frac{H(t_3)}{z_3} \right)^{n_3} \, \rangle_c.
\label{GOE3}
\end{eqnarray}
For GSE, we have to replace $N \rightarrow 2N$
in the right hand side of the definition (\ref{GOE3}).
First we discuss the GOE case.
Ignoring the contraction within the same trace,
a diagrams with $m_1, \ m_2$ and $m_3$ propagators is 
depicted in  fig 5 and fig 6. 
The number $n_a$ of operators in eq(\ref{GOE3}) 
can be expressed in terms of the number 
$m_a$ of propagators in a diagram in fig 5 or fig 6
\begin{equation}
n_1 = m_2 + m_3, \ \ \ n_2= m_3 + m_1, \ \ \ n_3= m_1 + m_2.
\end{equation} 
The summation can be calculated in terms of the 
number $m_1, \ m_2$ and $m_3$ of the propagators.
In the leading order of the $1/N$
expansion, there are $4 n_1 n_2 n_3$ diagrams with $m_1, \ m_2$ 
and $m_3$ propagators which consist of  $n_1 n_2 n_3$ 
untwisted diagrams in fig 5 and $3 n_1 n_2 n_3$ twisted diagrams in fig 6.
Therefore 
the three level correlator is calculated as 
\begin{eqnarray}
& & \sum_{n_1 = 1} ^\infty \sum_{n_2=1} ^\infty \sum_{n_3=1} ^\infty
\frac{1}{n_1 n_2 n_3} \langle \, \Tr \left( \frac{H(t_1)}{z_1} \right)^{n_1}
\Tr \left(\frac{H(t_2)}{z_2} \right)^{n_2} 
\Tr \left( \frac{H(t_3)}{z_3} \right)^{n_3} \, \rangle_c \nonumber \\
&\sim&
 \frac{4}{N} \sum_{m_1 =1} ^\infty \sum_{m_2=1} ^\infty
\sum_{m_3=1} ^\infty \left( \frac{K(t_2-t_3)}
{2 z_2 z_3} \right)^{m_1}
\left( \frac{K(t_3-t_1)}{2 z_3 z_1} \right)^{m_2} 
\left( \frac{K(t_1-t_2)}{2 z_1 z_2} \right)^{m_3} 
\nonumber \\
&+& \frac{4}{N} \left( \sum_{m_2=1} ^\infty  \sum_{m_3=1} ^\infty 
\left( \frac{K(t_3-t_1)}{2 z_3 z_1} \right)^{m_2}
\left( \frac{K(t_1-t_2)}{2 z_1 z_2} \right)^{m_3} 
+ 2 \ cyclic \ permutations \right).
\end{eqnarray}
To obtain the exact 3 level correlator, we have to take into account the 
vertex correction as well as the self energy correction 
to the one point function.
The vertex corrections can be done with 
the untwisted propagators as depicted in fig 7. 
Then the three level correlator becomes
\begin{equation}
G(z_1, \, z_2, \, z_3; \, t_1, \, t_2, \, t_3)
= - \frac{4} {N^4} 
\frac{\partial^3}{\partial z_1 \partial z_2 \partial z_3}
 F(z_1, \, z_2, \, z_3; \, t_1, \, t_2, \, t_3), 
\label{3level}
\end{equation}
where
\begin{eqnarray}
&& F(z_1, \, z_2, \, z_3; \, t_1, \, t_2, \, t_3)
\equiv \frac{ X_{23}}{1-X_{23}}{\frac{X_{31}}{1-X_{31}}}
{\frac{X_{12}}{1-X_{12}}}
\nonumber \\
&+&
{\frac{ X_{31}}{1-X_{31}}}\frac{X_{12}}{1-X_{12}}\frac{1}{1-X_{11}}
+{\frac{ X_{12}}{1-X_{12}}}\frac{X_{23}}{1-X_{23}}\frac{1}{1-X_{22}}
+\frac{ X_{23}}{1-X_{23}}{\frac{X_{31}}{1-X_{31}}}\frac{1}{1-X_{33}},
\label{3func}
\end{eqnarray}
with $X_{ab} \equiv K(t_a-t_b)/(2 z_a z_b)$ 
in this approximation. 
If we take into account the self-energy correction,
the free line $\delta_{ij}/z_a$ should be replaced by the 
dressed line $G_{ij}(z_a)$.  Therefore we have to use
\begin{equation}
X_{ab} = \frac{1}{2} K(t_a-t_b) G(z_a) G(z_b),
\end{equation}
where $G(z)$ is given in eq(\ref{GOE1}). 
The exact three level correlator in GOE is given by eq(\ref{3level})
with the function (\ref{3func}). This result agrees with that
obtained  by Verbaarschot, Weidenm{\"u}ller 
and Zirnbauer \cite{VWZ}.

Next we calculate the three level correlator in GSE.
All the contributing diagrams in this case are identical to those in the GOE
case. The contribution from the $n_1 n_2 n_3$ untwisted diagrams becomes
\begin{eqnarray} 
& & \frac{1}{N}\sum_{m_1 =1} ^\infty \sum_{m_2=1} ^\infty
\sum_{m_3=1} ^\infty \frac{1}{2} \left( \frac{2 K(t_2-t_3)}
{z_2 z_3} \right)^{m_1}
\left( \frac{2 K(t_3-t_1)}{ z_3 z_1} \right)^{m_2} 
\left( \frac{2 K(t_1-t_2)}{ z_1 z_2} \right)^{m_3}
\nonumber \\
&+& \frac{1}{N}\left( \sum_{m_2=1} ^\infty  \sum_{m_3=1} ^\infty  \frac{1}{2} 
\left( \frac{2 K(t_3-t_1)}{z_3 z_1} \right)^{m_2}
\left( \frac{2 K(t_1-t_2)}{z_1 z_2} \right)^{m_3} 
+ 2 \ cyclic \ permutations \right).
\end{eqnarray} 
Note that the number of loops with respect to the spin index
in an untwisted diagram with $m_1, m_2 $ and $m_3$ 
propagators shown in fig 5
$(\alpha= 1, \,  2)$ is always $m_1 + m_2 + m_3 - 1$. 
Contribution from remaining $3 n_1 n_2 n_3$ twisted diagrams 
in fig 6 is
\begin{eqnarray} 
 && \frac{3}{N}\sum_{m_1 =1} ^\infty \sum_{m_2=1} ^\infty
\sum_{m_3=1} ^\infty \frac{1}{2} \left( \frac{2 K(t_2-t_3)}
{z_2 z_3} \right)^{m_1}
\left( \frac{ K(t_3-t_1)}{ z_3 z_1} \right)^{m_2} 
(A^{m_2})^{\alpha \beta, \gamma \epsilon} 
\left( \frac{ K(t_1-t_2)}{ z_1 z_2} \right)^{m_3}
(A^{m_3})^{\epsilon \gamma, \beta \alpha}  
\nonumber \\
  &+& \frac{1}{N} \left( \sum_{m_2=1} ^\infty  \sum_{m_3=1} ^\infty 
 \left( \frac{K(t_3-t_1)}{z_3 z_1} \right)^{m_2} 
(A^{m_2})^{\alpha \alpha, \beta \gamma}
\left( \frac{K(t_1-t_2)}{z_1 z_2} \right)^{m_3} 
(A^{m_3})^{\gamma \beta, \epsilon \epsilon}
+ 2 \ cyclic \ permutations \right) \nonumber \\
&+& \frac{2}{N} \left( \sum_{m_2=1} ^\infty  \sum_{m_3=1} ^\infty 
 \left( \frac{K(t_3-t_1)}{z_3 z_1} \right)^{m_2} 
(A^{m_2})^{\alpha \alpha, \beta \beta}
\left( \frac{2 K(t_1-t_2)}{z_1 z_2} \right)^{m_3} 
+ 2 \ cyclic \ permutations \right).
\end{eqnarray} 
The contraction formula (\ref{contraction}) 
for the tensor $A^{\alpha \beta, \gamma \epsilon}$ 
enables us to calculate these contributions, and then 
the three level correlator in GSE is obtained as 
\begin{equation}
G(z_1, \, z_2, \, z_3; \, t_1, \, t_2, \, t_3)
= - \frac{1} {4 N^4} 
\frac{\partial^3}{\partial z_1 \partial z_2 \partial z_3}
F(z_1, \, z_2, \, z_3; \, t_1, \, t_2, \, t_3). 
\label{3level}
\end{equation} 
The function $F(z_1, \, z_2, \, z_3; \, t_1, \, t_2, \, t_3)$
is defined by eq(\ref{3func}) with the function $X_{ab}$ for GSE
$$
X_{ab} \equiv 2 K(t_a-t_b) G(z_a) G(z_b),
$$ 
with the one point Green function defined by eq(\ref{GSE1}).
Three level correlators in GOE and GSE obtained by the diagrammatic method 
are identical those calculated by the replica method and the 
functional method \cite{I}. \\

We have calculated the averaged one point Green's functions and the wide
connected two level and three level correlators in Gaussian orthogonal and
symplectic random matrix ensembles with time 
evolution by a diagrammatic  method.  All results are consistent with 
those obtained in all other methods \cite{VWZ,B,I,IMS} \\


We would like to thank S. Higuchi and H. Mukaida for helpful comments.


\newpage

\begin{figure}
  \begin{center}
    \leavevmode
    \epsfysize=2cm \epsfbox{fig1a.eps}\\(a)
  \end{center}
  \begin{center}
    \leavevmode
    \epsfysize=2cm \epsfbox{fig1b.eps}\\(b)
  \end{center}
\caption{The diagram of free propagator in the GOE case.
(a) an untwisted free propagator. (b) a twisted one.}
\end{figure}
\vspace{3cm}

\begin{figure}
  \begin{center}
   \leavevmode
   \epsfysize=5cm \epsfbox{fig2.eps}\\
  \end{center}
  \begin{center}
   \leavevmode
   \caption{The diagram of $g_n$.}
  \end{center}
\end{figure}

\begin{figure}
  \begin{center}
   \leavevmode
   \epsfysize=9.5cm \epsfbox{fig3a.eps}
   \hspace{2cm}
   \epsfysize=9.5cm \epsfbox{fig3b.eps}
  \end{center}
\hspace{3.3cm}(a) \hspace{8.2cm} (b)\\
\caption{(a) An untwisted planer diagram in the $n$th order 
with the definition of $n$ untwisted contraction parts.
(b) A twisted one with definition of $n$ twisted contraction parts.
In the case of the GOE, the spinor indices 
($\alpha_{1}$,$\beta_{1}$,\dots) should be neglected.}
\end{figure}
\vspace{1cm}

\begin{figure}
  \begin{center}
   \leavevmode
   \epsfysize=2cm \epsfbox{fig4a.eps}\\(a)
  \end{center}
  \begin{center}
   \leavevmode
   \epsfysize=2cm
   \epsfbox{fig4b.eps}\\(b)
  \end{center}
\caption{The diagram of free propagator in the GSE case.
(a) an untwisted free propagator. (b) a twisted one.}
\end{figure}

\begin{figure}
  \begin{center}
   \leavevmode
   \epsfysize=5cm\epsfbox{fig5a.eps}
    \hspace{1.5cm}
   \epsfysize=3cm \epsfbox{fig5b.eps}
   \hspace{-0.45cm}(a) \hspace{6.62cm} (b)\\
  \end{center}
\caption{An untwisted diagram with $m_1$,$m_2$ and $m_3$ propagators. 
(a) $m_1$,$m_2$ and $m_3$${\neq}0$ case. 
(b) $m_1=0$, $m_2$ and $m_3$${\neq}0$ case.}
\end{figure}
\vspace{1cm}

\begin{figure}
  \begin{center}
   \leavevmode
   \epsfysize=5cm \epsfbox{fig6a.eps}
    \hspace{1.5cm}
   \epsfysize=3cm \epsfbox{fig6b.eps}
   \hspace{-0.45cm}(a) \hspace{6.62cm} (b)\\
  \end{center}
  \begin{center}
   \leavevmode
   \epsfysize=3cm \epsfbox{fig6c.eps}
    \hspace{2.03cm}
   \epsfysize=3cm \epsfbox{fig6d.eps}
   \hspace{0.1cm}(c) \hspace{7.6cm} (d)\\
  \end{center}
\caption{A twisted diagram with $m_1$,$m_2$ and $m_3$ propagators.
(a) $m_1$,$m_2$ and $m_3$${\neq}0$ case.
(b),(c) and (d) $m_1=0$, $m_2$ and $m_3$${\neq}0$ case.}
\end{figure}

\begin{figure}
  \begin{center}
   \leavevmode
   \epsfysize=3.7cm \epsfbox{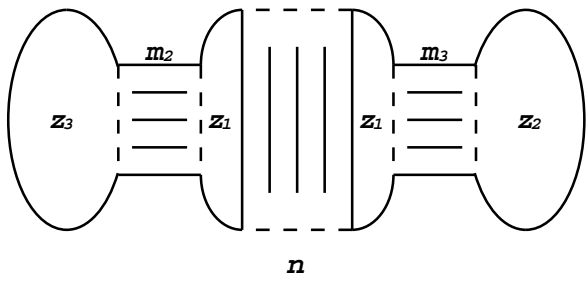}
    \hspace{2cm}
   \epsfysize=3.7cm \epsfbox{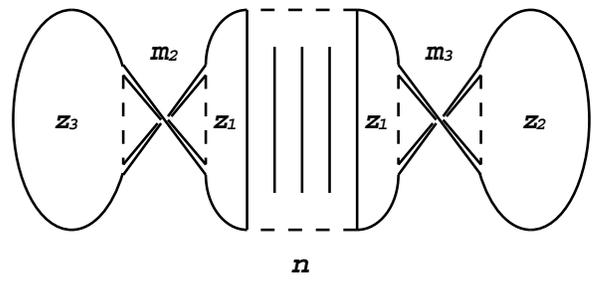}
   \hspace{0.1cm}(a) \hspace{9.05cm} (b)\\
  \end{center}
  \begin{center}
   \leavevmode
   \epsfysize=3.7cm \epsfbox{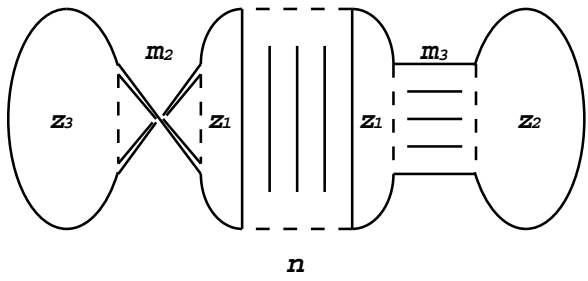}
    \hspace{2cm}
   \epsfysize=3.7cm \epsfbox{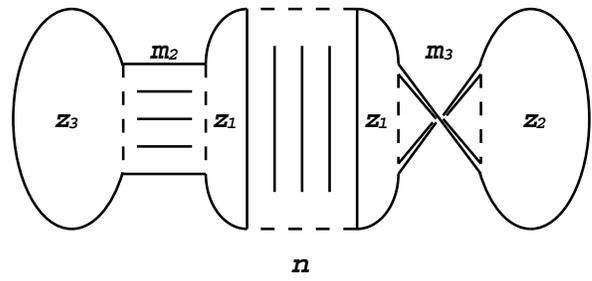}
   \hspace{0.1cm}(c) \hspace{9.05cm} (d)\\
  \end{center}
\caption{Diagrams which require vertex correction. $(n {\geq} 1)$}
\end{figure}

\end{document}